# The "revolution" in physics of the early Nineteenth century revisited in the context of science-and-society interaction

**Angelo Baracca**

Department of Physics, University of Florence

**Asbstract**

The radical changes in the concepts and approach in Physics at the turn of the Nineteenth century were so deep, that is acknowledged as a "revolution". However, in 1970 Thomas Kuhn's careful reconstruction of the researches on the black body problem, the concept itself of the "revolution" seemed to vanish in his diluted discussion of every details.

In the present paper, after an examination of the limitations of Kuhn's response to his critics, I put forward the idea – although it is not new – that these changes in Physics cannot be reduced to a "point-like" event, but happened instead through multiple successive (and even contradictory) changes in the course of decades. Such as the "old" quantum hypothesis, wave mechanics, "orthodox" quantum mechanics. In fact, the innovative perspectives started in the 1980s have been considered as a "third quantum revolution".

My basic argument is that these changes, in order to be really understood, must be interpreted not as mere specific changes in Physics, but framed in the context of the deep social, cultural, and economic changes during those turbulent years. The main steps are outlined.

The concept that at the beginning of the Twentieth century physics underwent a deep "revolution" in its foundations and basic concepts is largely acknowledged both in the scientific and the philosophical-historical communities. Even the basic textbooks of physics adopt this concept, although they usually present a stereotyped *a posteriori* "rational" reconstruction, which misrepresents the true genesis and development of the new concepts. On the other side, an enormous quantity of detailed and rigorous studies has been produced on the subject in the past decades, which has brought a deep advancement in the reconstruction and interpretation of the scientific and epistemological contributions of the main scientists. Nevertheless, such a tremendous quantity of information has not resolved, rather has deepened, the controversies on the interpretation and the exact identification of this "revolution".

The concept of *scientific revolutions* was introduced in a famous book by Thomas Kuhn (1970), in which the concept of *paradigm* was also proposed. However, when Kuhn successively presented the most careful and detailed reconstruction of the researches on the black body problem at the turn of the Nineteenth century (Kuhn 1978), the concept itself of the "revolution" seemed to vanish in his diluted discussion of every details. This remark was raised by several scholars (Klein, Shimony, and Pinch 1979; Needell 1980; Galison 1981; Darrigol 1991, 1992; Jost 1995; Gearhart 2002), and Kuhn replied proposing in practice a "point localization" of the "revolution" (Kuhn 1982, 1984). In the development of the controversy, a firm point is the analysis by Büttner, Renn and Schemmel (2003), based on Einstein's correspondence with his fiancée Milena Maric, to which we refer for more details (not to speak of previous papers, and the whole critical edition of Einstein's papers by Renn).

In the present paper we argue that a complementary point of view and reconstruction can greatly help in getting a deeper, or wider, insight into the process of scientific development and innovation. We will resume the main arguments, specifying the basic considerations.[1] We will start with a critical consideration of Kuhn's concepts, and the concept itself of "scientific revolution".

**1. On the concept of "scientific revolution"**

In our opinion, several criticisms can be moved to the concept of "scientific revolution" as proposed by Kuhn.

**1)** Kuhn´s concept of "scientific revolution" is *too generic*, in some sense even abstract. Although at first sight the change in *paradigms* appears quite clear (e.g., the turn from classical to quantum theory), when a deeper analysis is developed it appears not so easy to identify and pinpoint the change, its nature and substance: it seems just what emerges in the controversies on the quantum "revolution". The point is that there is no clear-cut criterion for identifying the change. On the one side, the evaluation of the change of paradigm strongly depends on personal appreciation. On the other side, when the reconstruction of the developments of the physical concepts is carried out in great details, resorting to private correspondence or documents of the scientists, even psychological aspects enter into play (if not psychoanalytical ones, as for instance in the case of the relationship between Pauli and Jung).

Obviously, the rigorous reconstruction of the evolution of science in every details, having recourse to every document that can throw light into the process of scientific elaboration and creation, brings fundamental contributions for understanding the development of science. Our point is that this analysis in deep should be performed having some general criterion for placing and interpreting the specific developments. Otherwise, what does really represent every change in the conceptions of a scientist for the general evolution of science? How many, even deeply creative, changes in scientific conceptions did happen, and can be singled out in the thought of scientists, which had no concrete influence in the development of science? What determines that an innovative idea results in a change in science? Not all the discarded ideas were wrong; sometimes they were only premature: but why?

**2)** Kuhn´s concept of "scientific revolution" is at the same time *too restricted*. Scientific changes do not have a unique, "point-like" origin and meaning, uniquely determining all subsequent scientific developments. As an example, when *models* were introduced in physics and chemistry after mid Nineteenth century, this happened in several ways, and with successive changes, which were not merely implicit in the early idea. Actually, *models* had been proposed even earlier, but had not been generally accepted by the scientific community as general tools: which conditions were necessary for their general acceptance and adoption? Moreover, during the 1850s *models* were introduced *independently* in physics and in chemistry. Finally, these early models evolved in substantial, and unforeseen, ways in the subsequent decades, originating general scientific *theories*. This "revolution" was instead a *process*.

In a similar way, the Twentieth century "revolution" in physics did not originate in a unique hypothesis, but involved different disciplines, and underwent a long development, through successive and unforeseen changes. The theory of relativity was *a* "revolution", not implicit in, but connected with, the "quantum" one. And, in this respect, why the previous relevant contributions of Lorentz and Poincare did not resort in a new theory? The "revolution" in physics cannot be reduced, in time and space, to some new concept introduced in 1900 or in 1905, or somewhere in between (be it due to Planck or Einstein). It was instead a *process*, a long and very complex one, which covered a quarter of a century. The elaboration of orthodox quantum mechanics at mid 1920s was not implicit in the 1900-revolution, it rather introduced a really new paradigm. Otherwise, one could not understand why Einstein, who was (at least one of) the main authors of the Twentieth century "revolution", was so hostile towards the successive evolution of the theory in the 1920s. In our view (as we will clarify in the following), both Planck and Einstein introduced in the early 1900s fundamental, and independent, pieces of changes which contributed to the early quantum framework: and which were overcome by (substantially incompatible) transformations of mid 1920s.

---

[1] This approach and interpretation was proposed in an early essay (Baracca, Ruffo and Russo 1979), in Italian, and resumed and extended for this specific subject in 2005 (Baracca 2005).

**3)** On the general concept of "revolution", one should consider that every kind of, however radical, social change is not a sudden and localized historical event, but rather a *process*. The French Revolution cannot be restricted to the events of 1789, as if the subsequent transformations were implicit in the first changes: it was instead a long lasting process, starting off new and original innovations, whose early seed was moreover not suddenly thrown in 1789, but long prepared not only by previous transformations in France, but also by foreign influences (like the Enlightenment movement directly inspired to the English example). Similarly, the Bolshevik Revolution was not an event restricted to October of 1917, which could not prescind from previous movements and attempts, and threw a seed that was to produce *new* and *unpredictable* transformations.

**4)** Production of science is a social activity, deeply intertwined with the whole social environment, its contradictions, cultural currents and traditions, the specific role of scientists and the organization of scientific activity (with national peculiarities), up to the hard economic and technological demands, choices and transformations, including the advancements in experimental techniques. It is a fact that when deep changes happen in science, they usually go along with deep changes not only in the social role of scientists and in the way they conceive and carry out their profession (which is not a purely *scientific* aspect), but also with deep changes in culture, in society, in technology, in economy. In the majority of cases one finds out common, or parallel, features in all (or in several of) these areas. The consideration of these connections, and of the more general social context, can throw a deeper light into each of these areas, and in particular can help in judging and placing the conditions in which new scientific concepts and hypotheses arise, and allow them to overcome previous conceptions and methods, and to grow into new scientific "paradigms" and methodological frameworks.

An enlightening comparison is in our opinion D. S. Landes' reconstruction of the technological changes and industrial development in Europe (Landes 1965): it represents a true *social history of technology*, in which those technological changes that in other, even more detailed, histories of technology appear as simply due to individual inventiveness or talent or cumulative skills, acquire a deeper meaning and justification in connection with the social and economic changes and requirements. The industrial revolutions were *processes*, which did not originate in a single innovation, but were molded in the course of decades under the incentive of new economic and technical needs, through the multiple introduction of innovations and their successive improvements, which often introduced substantial innovations (for instance, for steam power was the "crucial" invention Savery´s engine of 1698, Newcomen´s engine of 1712, or Watt's invention of 1769?). Why should things go in a different way for scientific innovations?

Not differently, the above mentioned *process* of the *introduction* and *development* of *models* in physics and chemistry in the second half of the 19th century can be most properly understood and interpreted making substantial reference to the social and economic conditions and requirements for changes in economy, in industry, in science, in education, in culture (Baracca, Ruffo and Russo 1979).

We are fully aware of the danger of confusing simple *associations*, and *analogies* in different fields for a *causal* connection. We maintain, however, that a definite historical situation stimulates, or generates, innovations (both practical and intellectual), conceptual frameworks, new trends and concepts, which show concrete similarities, much deeper than what we can discuss in the present concise survey; and that – even apart from direct connections (which nevertheless in many cases can be demonstrated) – they provide a more complete, understandable and convincing reconstruction of the evolution of science and its deep changes.

**5)** By the way, let us briefly mention one more advantage that we see in the approach we are proposing, referred to the popularization of science. This is an extremely demanding problem, since one cannot obviously pretend that common people masters the formal content of modern science. Even the students of physics meet some difficulty against modern theories only on the basis of formalisms. We deem that reference to the general historical context, the technological changes, the cultural and intellectual currents, the changing social role of scientists, the material transformations and requirements posed by the social and economic environment, can greatly help common people to catch the reasons, and also the specific basic features, of the scientific approaches to natural phenomena and their physical representations.

On the basis of the previous considerations, let us briefly summarize how the revolution (or revolutions?) in physics (and not only) in the first quarter of the Twentieth century can be interpreted. In spite of our previous criticism, we will continue to use the word "revolution", however we consider more appropriate and unambiguous to use in many (even fundamental) aspects such words as *turns*, or *turning points*, in the reconstruction of a whole and complex *process* of scientific change.

## 2. Nineteenth century prodromes and anticipations of the Twentieth century scientific revolution in physics

*2.1. The contradictions of the physicists at the end of the Nineteenth century*

In the last decades of the 19th century the situation in physics appears deeply contradictory. The "paradoxes" which plagued the big 19th century theories were actually only the epi-phenomenon of a deep contradiction. Indeed, the *models* that had begun to be introduced in the 1850s had evolved in the subsequent decades into a general *mechanistic world view*, which was the advanced frontier of physics at the turn of the century: strong opposition against this program and methodological framework came from contrasting positions, which refused the atomic-molecular model and referred to a substantially phenomenologi cal attitude (Mach's Empirical criticism, and Ostwald's Energetics). It was to become evident that the "paradoxes" ("ether wind" for electromagnetism, "irreversibility" and "recurrence" for kinetic theory) originated only from a forced mechanistic interpretation of these physical theories, since both were to be embodied respectively in the theory of special relativity and in statistical mechanics.

After all, the basic contradiction of 19[th] century physics can be identified in the fact that, since the adoption of *models* had been acknowledged as a powerful probe to achieve new scientific knowledge (as for instance the transcendental prediction of electromagnetic waves), the *limitation to mechanical models* appears *a posteriori* as contradictory and absurd. There were for that *external general conditions*, which prevented the introduction and consideration of new conceptions.

One can remark that, parallel to the mechanistic view, an electromagnetic one was also developed, but it shared with the mechanistic one a basic *reductionist* attitude: in both the physical properties were *built-up* from the underlying substrate and elementary components, be their of a mechanical or electromagnetic nature. Lorentz' "theory of electrons" led to results physically equivalent to the subsequent theory of special relativity: the main, but substantial, difference was that those results which in Lorentz' theory required complex calculations from the set of Maxwell's equations plus Lorentz' law, resulted instead as straightforward consequences of the postulates of special relativity. In fact, although Lorentz and Poincaré introduced basic concepts of the subsequent theory of special relativity, neither of them went as far as to formulate a new physical theory.

We would anticipate that the new physical theories which were proposed in the early 1900s inherited aspects of both the reductionist and the phenomenological attitudes, but inserted them into a completely new kind of physical theory and concept of natural reality (for instance, Einstein declared his debt towards Mach's conceptions, and it is well known the critical role of the observer in special relativity, but one could not absolutely reduce his attitude to a positivistic one, and his theory to a Nineteenth century one).

*2.2. The mismatch between physics and chemistry at the end of the 19[th] century, and the peculiarities of technological and scientific innovation in different fields.*

The above mentioned contradiction stands even more out against the radically diverging attitude adopted by chemists (Baracca, Ruffo and Russo 1979, Baracca 2005). Physics and chemistry had shared since the 1850s the choice of relying on *models*, which had allowed a big leap, leading both disciplines to acquire a powerful predicting and creative power, and meeting the challenges posed by a deeply renewed wave of industrialization and technical innovation. The latter was a *process*, which had a long period of preparation during the 1850s and 1860s, and took off in the last decades of the century, rightly called Second Industrial Revolution (Landes 1965). This take off was centred in Germany and occurred in several fields, mainly in the chemical and the electric industries. In the chemical sector competition was very strong, professional chemists and engineers were deeply involved in the development of the new, highly technological, chemical industry (BASF, Bayer, Hoechst, etc.), holding roles of big responsibilities, since the management of the firms were grounded on laboratories and advanced research (even through organic collaborations with universities and polytechnics).

The main problems chemists had to solve concerned the determination of equilibrium conditions for chemical reactions (above all for the new and extremely complex organic ones). Such a problem could not even be tackled on the basis of a reductionist-kinetic approach. The chemists were thus induced, without controversies, to adopt *a new scientific approach, based on the synthetic and universal laws of thermodynamics*. So, in the last decades of the 19[th] century they developed the field of *theoretical thermodynamics*, based on the concept of "free energy": Willard Gibbs published in 1876 a systematic paper on the equilibrium of heterogeneous systems, which was quite formal and complicated, while a series of more practical criteria and laws were established which allowed to manage the problem of chemical equilibria.

In some sense one could say that the chemists "replaced" thermodynamics for mechanics as the basic framework. This choice, under the pressure of challenging technical scientific demands, anticipated the methodological change that was to be at the basis of the innovations introduced at the beginning of 1900s by the physicists, when the challenges of new technical requirements and new discoveries began to break up the reassuring reductionist world view (characteristic spectral lines, X-rays, cathode rays, discovery of the electron, radioactivity). In the electrotechnical sector one of the main challenges concerned electromagnetic radiation. Big technical improvements in the detectors of electromagnetic waves law led to the final measures on cavity radiation at the *Technische Physikalische Reichsanstalt* in Berlin in the year 1900.

*2.3. Planck's peculiar role and research at the end of the Nineteenth century*

Max Planck, for all his attitudes, positions and actions all along his life and career, could hardly be classified as an "innovative" personality (see e.g. Heilbron 1986). Actually, his position in physics at the end of the century was peculiar, and isolated in the scientific debate. He did not adhere to the mechanistic approach, but opposed Mach´s and Ostwald`s views, and moreover also Boltzmann's statistical interpretation of the second law (see e.g. Badino 2000), that he considered instead as an *absolute* law; he therefore did not even share the atomic model of matter. *This can hardly be qualified as an advanced, or "revolutionary", although original, attitude. Is it conceivable therefore that, shortly after, he suddenly turned into a "revolutionary"?*[2]

Nevertheless, we would assert that it was just Planck´s reliance on "absolute" thermodynamics that allowed him to introduce an innovative approach to the black-body problem

During the 1880s and 1890s Planck elaborated a *thermodynamic* approach to the black-body problem (see the careful reconstruction by Badino 2014; interesting also Gearhart 2002), based on the entropy *of a single resonator* (in fact, a concept incompatible with the statistical nature of entropy): in fact, he adopted as a basic *tool* the second derivative of the entropy of a resonator with respect to its energy.

On this original basis, Planck (1900a) finally succeeded in "connecting" Wien's heuristic, widely accepted, law for the frequency distribution of cavity radiation to the simplest parametrization, i.e. a direct proportionality of the second derivative of the entropy to the energy of the resonator (see e.g. Baracca 2005, Appendix, for technical details).

*2.4. The small world of physics at the turn of the century, amid deep intellectual unrest*

We deem it relevant to take into account the smallness of the world of physics at the turn of the century, as compared to the dimension and dynamism of modern research. The number of academic physicists in each whole country was comparable to that of the university teaching staff of a single Department of Physics in a developed country at present. Some numbers of academic physicists (Senior, Junior, and Privat-dozents, and [in square parenthesis] the total numbers adding Assistants and Research affiliates), in 1900 reported by Forman, Heilbron and Weart (1975, 12) are (note the relatively high number in Germany): Austria-Hungary, 48 [79]; U.K., 76 [144]; France, 53 [145]; Germany, 103 [235]; Italy, 43 [73]; United States, 99 [195]. The number of scientific journals was very limited, and the new ideas were rapidly shared by the entire physical community.

---

[2] We rather agree with Krag's definition as a "reluctant" or "conservative revolutionary" (Kragh 2000).

The whole intellectual environment was shaken by deep restlessness and innovative ferments that upset the cultural and methodological horizon of the 19<sup>th</sup> century. But it was too a very limited environment. As Janik and Toulmin (1973, p. 92) describe the intellectual environment in Vienna:

> «It is not easy today, especially for a younger American, to recognize just how small and tightly knit were the cultural circles of the Habsburg monarchy. [...] Mass education makes it difficult, likewise, to conceive of a country in which there was only one real university, and that contained pretty much in one single building; […] Thus it comes as a slight shock to discover that Anton Bruckner gave piano lessons to Ludwig Boltzmann; that Gustav Mahler would bring his psychological problems to Dr. Freud; that Breuer was Brentano's physician, that the young Freud fought a duel with the young Viktor Adler, who had attended the same high school as both the last of the Habsburgs, Charles I, and Arthur Seyssinquart, later the Nazi Commissioner of Holland; and that Adler himself, like Schnitzler and Freud, had been an assistant in Meynert's clinic. In short, in late Habsburg Vienna, any of the city's cultural leaders could make the acquaintance of any other without difficulty, and many of them were in fact close friends despite working in quite distinct fields of art, thought and public affairs.»

In that same limited environment acted in those years persons like Ernst Mach, Schönberg, Wittgenstein, Klimt, Kokoschka.

Behind all that, the 19$^{th}$ century political, economic and social order was at the brink of its disintegration (*ibid.*, 63):

> «A layer of waltz and whipped cream coverrd at the surface a desperate society … any proportion between semblance and reality had disappeared.»

### 3. The multiple take off of the (first phase of the) "revolution" in the early Twentieth century: the breaking of thermodynamics into physics

In our opinion, Planck and Einstein brought different and complementary (and multiple) contributions to the deep changes in physics in the early 1900s: which started a *process*, and were in turn to be quite radically overcome in the way of the completion of the *process* with the formulation of orthodox quantum mechanics.

*3.1. The two 1900 Planck's papers: a first turning point in the* process *of innovation in physics, a non-mechanical thermodynamic approach.*

In fact, we think that the *fundamental innovation* carried out by Planck in the year 1900 (already adopted by him, as we have remarked, in the previous paper, at variance with the dominating currents in physics) was the *adoption in physics of the first non-reductionist model*, properly a *thermodynamic* one.[3]

In the *first one of the two 1900 Planck's papers on black body spectrum* (1900b), searching for the correct law for the newly obtained experimental results, he "simply" extended his previous parameterization (a procedure which is widely adopted by physicists), assuming a proportionality of the second derivative of the resonator entropy to a linear combination of the first and second powers of the resonator energy.[4]

---

[3] Martin Klein (1962, 1963, 1964, 1966, 1967, 1982) had already remarked the role of thermodynamics in the early papers of Planck and Einstein in the early.

[4] By the way, let us insist on how the usual reconstruction of these developments in the textbooks of physics is completely false, referring to the physical contradictions of the so-called "Rayleigh formula" based on equipartition, which left Planck the only choice of a statistical approach *a la* Boltzmann. In fact, not only in the year 1900 Rayleigh did not propose the so-called "Rayleigh formula", but simply a

We would remark that this step was a substantial innovation in physics (not for chemistry, as we have discussed, although with different purposes and features), but it had at the same time a "conservative" nature, since it was based on an absolute interpretation of the second law, and a refusal of the atomic concepts.[5]

Regarding his second (and generally acknowledged) paper on his new law (Planck 1900c), we agree with those who deny that it introduced the "revolutionary" quantum concept.[6] We base this assertion on several considerations.

In the first place, Planck definitively does not consider the "energy element" $h\nu$ as a physical quantity: in fact he explicitly writes:

> «When the ratio [between the energy of the oscillators of a given frequency $\nu$ and the "finite part", or "energy element", $h\nu$] is not an integer, we take for it the nearest integer.»

Planck considered for a long time this as a mathematical hypothesis, an artefact (Kragh 2000): the procedure of "discretization" of a continuum problem was quite common at that time. In fact, for many years thereafter Planck was extraordinarily reticent in ascribing any physical interpretation of these energy elements and of his new constant h (Planck's Nobel Lecture, 1920, p. 108-109 of the transl.; Planck 1949, p. 43-45; Planck's correspondence discussed by Gearhart 2002).

In the second place, how can one interpret the claim that in this second paper Planck – so abruptly! – adopted Boltzmann's statistical concepts, that he had strongly opposed till then? In our opinion things went quite differently. In his Nobel Lecture, Planck qualified the procedure in his second paper as an «act of desperation». It is quite different to adopt instrumentally Boltzmann's "formalism" (moreover partially, as we argue) for the purpose of obtaining a positive result, from adopting Boltzmann's "theory": as he declared much later:

> «It was a purely formal hypothesis, and I certainly did not devote much attention to it: the only thing that interested me, at any cost, was to arrive at a positive result.» (Planck 1931)

The more so, since "Boltzmann's theory" was the *inseparable union* of the law $S = k \ln W$ and the expression for $W = N!/\prod n_i!$. It is well known that Planck, instead, wrote down for $W$ a completely different expression:[7] one which anticipated of 24 years the Bose-Einstein statistics. Planck himself acknowledged that:

---

formal change in Wien's heuristic law (see the detailed reconstruction of Kuhn, p. 144-52); moreover the so-called "Rayleigh formula" was *never* proposed as the physical distribution of cavity radiation (Avila and Baracca 2006). In 1904, when Planck's law had been firmly experimentally confirmed, Rayleigh considered the above mentioned formula overturning the question: why equipartition does not apply to cavity radiation?

[5] We deem however interesting to remark that in a statistical approach the second derivative of the entropy of the resonators is directly related to their energy fluctuations (or of the correspondent natural modes of cavity radiation). So that, curiously enough, had Planck accepted statistical mechanics (which in fact was developed by Einstein in 1902-1904), the parametrization that he introduced in 1900 in order to reproduce the new results for the spectrum would have implied the wave-particle duality! Which in fact was obtained by Einstein in 1909 on the basis of fluctuations (see Baracca 2005, Appendix, for details).

[6] By the way, we disagree on this point with the, however careful and rigorous, reconstruction of Planck´s research by Badino (2012): in fact, he completely disregards the first Planck paper (1900b) in which he derived his complete law, and this leads in our opinion to a misinterpretation of Planck´s 1900c paper.

[7] As Gearhart remarks: «This expression … gives the total number of complexions, and stands in sharp contrast to Boltzmann's procedure: Boltzmann had picked out a subset of complexions corresponding to a given macroscopic state, and by a maximization procedure found the particular subset corresponding to the most probable (and hence, equilibrium) state» (Gearhart 2002, p. 202).

> «In my opinion, this hypothesis essentially corresponds to *a definition of the probability W*» [italics are our] (Planck 1901)

Only in about 1912 Planck accepted the truly statistical nature of the second law (Kragh 2000).

One more argument is that, adopting Boltzmann's theory, Planck should have calculated the maximum value of *W* in order to get the frequency distribution at equilibrium: but actually he was satisfied since he directly got that «very simple logarithmic expression» of his previous paper that was equivalent to his new law. The same fact justifies, in our opinion, why, at the end of the procedure, Planck did not take the limit for *hv—>0*.

*3.2. Einstein's papers, 1902-1905: the decisive turns of a multifaceted "revolution".*

Our criticism of the restrictive viewpoint that underlies the concept of the Twentieth century "revolution" in physics is reflected in circumscribing the changes to the birth of the quantum concepts. Actually, the deep innovations (the whole "revolution") involved *at least three* basic physical disciplines, including statistical mechanics and the theory of relativity: the physical nature of the changes introduced in the latter two disciplines had nothing to do with the introduction of the quantum, although they stemmed from an analogous turn in the methodological attitude, i.e. the overcoming of a reductionist approach, and specifically the resort to (statistical) thermodynamic reasoning.

*3.3. The foundational and innovative role of statistical thermodynamics*

In the first place, *statistical mechanics* (it should be most properly called *statistical thermodynamics*) was not a straightforward development of kinetic theory, even substantially enriched by probability concepts. Actually, Boltzmann had introduced practically all the basic concepts and formalisms (even statistical ensembles), but he did not formulate, not even fully acknowledge in 1902-1904, a general statistical theory (Einstein 1902, 1903, 1904). Willard Gibbs (1902) *formally* introduced statistical methods in mechanics as a sort of "rational thermodynamics", without even the claim that they had to reproduce the experimental properties, Einstein's papers in 1902-1904 posed the foundation of a *new physical theory*. In which the *basic role of thermodynamics in place of mechanics* is even self-evident: thermodynamic functions, that in kinetic theory were obtained through lengthy and complex calculations of the collisions between atoms, come out as normalization "constants" (in the phase space, but "functions" of the macroscopic thermodynamic variables) of an *abstract probability function*. In fact, the expression of this abstract probability in the phase space for a specific macroscopic thermodynamic condition (e.g. microcanonical, or canonical ensemble) is such that the normalization functions are exactly the proper thermodynamic functions (entropy, or Helmholtz free energy, respectively). In this framework, the "paradoxes" that had plagued the debate in physics simply vanished.[8] In fact, dealing with a system of bodies, one can make two complementary choices, excluding each other: *either* one can determine the exact positions and velocities of all of them, and then *must* apply Hamilton's equations and determine their trajectories; *or*, for any reason, one has only an approximate knowledge of positions and velocities, and then *must* adopt the statistical formalism (see for a technical exposition Baracca 2005, Appendix).

---

[8] One could remark that the crisis of classical physics was implicit (latent) in the statistical meaning of thermodynamics, and the asymmetry of time: one had "simply" to invert the roles of the "paradoxes" and their interpretation.

By the way, it seems appropriate to remark the generality of such direct reliance on thermodynamic laws. The latter are in fact independent from the nature and behaviour of the underlying microscopic substrate. In fact, the later formulation of *quantum statistical thermodynamics* would rely on the same identical formalism of statistical ensembles, substituting integrals in the phase space for discrete sums on microscopic quantum states. Such generality stemmed from Einstein's concept of thermodynamic laws as universally valid: in their intrinsic statistical nature, not in Planck's absolute concept. This reinforces our thesis that Planck cannot be acknowledged as the author of a "revolution" *tout court*, although he performed a first fundamental step.

Einstein's physical formulation of statistical thermodynamics led him to acknowledge the basic role of *fluctuations*: a tool which opened the way to his consideration of the Brownian motion in one of the 1905 papers (Einstein 1905b), and to the skilful experiments with which Perrin finally demonstrated the physical reality of atoms, opening moreover a completely field which was to have huge developments, that of *stochastic processes*.

*3.4. The light-quantum as a real physical entity.*

For our purpose, our basic remark is that Einstein (1905a) ascribed, beyond any doubt, a *physical reality to the quantum* [9]. Einstein´s approach confirms his methodological reliance on (statistical) thermodynamics, since the corpuscular light quantum hypothesis (against the overwhelming evidence accumulated of the wave nature of light) was "legitimated" by the formal analogy of the logarithmic dependence for entropy variation in an isothermal volume change in a gas and in cavity radiation.

By the way, these fundamental papers, when they appeared were not immediately acknowledged in the scientific environment: the first official discussion happened at the 1011 Solvay congress, but the light-quantum was not accepted until the 1920s, after the experiments by Compton.

*3.5. One further face of the "revolution", the theory of relativity*

The third of Einstein´s papers on *special relativity* (1905c) inaugurated one further face of the turns in physics, starting a long path which was to upturn our general conception of physical reality, even in popular perception. Einstein overcame with a single step the previous lengthy debate on the "ether wind", simply attributing a physical meaning to "Lorentz transformations" and assuming electromagnetism as the intrinsically covariant theory, while he reformulated Newtonian dynamics. The $19^{th}$ century (Newtonian) mechanistic worldview was overturned by one step, while a fourth paper (1905d) revolutionized the concepts of mass and energy.

*3.6. More pieces of the puzzle.*

Other deep changes, which were gradually taking shape, substantially contributed to the radical transformation of the concept of physical reality. One of them was the ascertainment of the reality of atoms, and of their internal structure, which slowly developed in a completely independent way of the new quantum concepts and problems (until 1913 the atom models were aimed to explain chemical properties). For instance, such a basic concept as the *atomic number* was not known until the results of Moseley of 1913.

---

[9] By the way, we refer to the Renn´s 1993 analysis, reconsidered in Büttner, Renn and Schemmel (2003), for the reconstruction of Einsten's choice of limiting himself in this first paper to the old Wien's formula for the spectrum, almost "disregarding" Planck´s law: it seems even more significant to see how the "analogy" with the Maxwell-Boltzmann distribution resulted in Einstein's reasoning into a new concept.

We would remark also the contribution of the improvements of the experimental techniques in changing the concept of physical reality. As an example, the dimensions of the *cosmos* went through subsequent big steps thanks to new observing and measuring equipment.

### The new rational and realistic – non-mechanistic, non-reductionist – worldview, 1900-1910, as the basic requirement for a physical theory

Those discussed above are in our opinion the radical innovations that all together set up the (early) radical changes in physical theories at the beginning of the 20$^{th}$ century. We want to remark that a completely new world view emerged from those changes, completely different by the 19$^{th}$ century one. A world view that Einstein considered, also in his subsequent activity, as a basic requisite to which a *physical theory* (even *physical reality*) must satisfy. In his view, a physical theory must describe a physical system and its evolution *in space and time*. One can recognize such a "realistic" requisite at the basis of Einstein's later criticism to Orthodox Quantum Mechanics and the role of the observer, such as the Einsten, Podolski and Rosen paradox. In fact, Eistein coherently developed his research program, widely based on (statistical) thermodynamics, which brought outstanding results, such as stimulated emission of radiation, Bose-Einstein statistics and Einstein condensation, the quantum theory of the ideal gas, and so on. As a matter of fact, Einstein, de Broglie, Schrödinger found themselves increasingly isolated, since the scientific community chose a pragmatic and formal approach. Resuming the comparison with Planck, one can remark that the latter explicitly expressed his scepticism towards Einstein's method and the new statistics (Planck 1925a, 1925b), and insisted in a formal, unjustified division by N! In order to avoid Gibbs' paradox.

It is evident that Quantum Mechanics has subsequently abandoned this requisite of a space-time description. Even the "revolution" in 20$^{th}$ century atomic (quantum) physics was therefore far from completed by the new approaches, besides the epistemological turn, developed at the beginning of the century by Planck and Einstein, whoever had been the author of the early turn (we discussed how both contributed in different forms to this *process*). In this there was a further difference in the subsequent evolution of the Theory of Relativity which, being developed mainly by Einstein, did not break, rather considerably enriched, that realistic world view (for these aspects we refer to Renn 2006, 2007).

What turns happened therefore after these early years? What had changed in the 1920s? Again, we think that the substance of the process can be appreciated in a rational and deeper form making reference to the most general historical environment. Even an outline of these further developments goes beyond the purposes of the present paper. We will merely mention the basic considerations.

### The collapse of the Nineteenth century world order, and of the reliance on a rational world view

It was not only the ephemeral glamour of the *Belle Époque* to be suddenly overwhelmed by the clangour of the war. It is sufficient to compare two maps of pre- and post-World War I Europe in order to appreciate the deepness of the collapse of a whole world order: the big magnificent, however decadent (how Robert Musil's humour masterfully depicted what he named "Kakania"), Austro-Hungarian empire that occupied the centre of the continent, had suddenly dissolved; the same for the Ottoman empire; while the other big empire at the Eastern side had radically changed its political colour.

The deep innovations introduced by the Second Industrial Revolution were finally to irretrievably clash with the Nineteenth century economic and political order, accelerating the contradictions and competition between the rapidly growing industrial societies.

However, the collapse of that world did not result in the immediate creation of a new order, but opened the way to a long and troubled historical phase of deep instability, which lasted, with dramatic developments, during the whole inter-war period (actually, Hobsbawm interprets these two decades as an entire war period).

The subversion of the world order had to reflect in the world view, in mentality, in the cultural and philosophical foundations, as well as in the role of scientists and their specific production.[10] As Arnold Sommerfeld eloquently remarked in one of the most prestigious of the South German monthlies:

> «The belief in a rational world order was shaken by the way the war ended and the peace dictated; consequently one seeks salvation in an irrational world order.» (Sommerfeld, 1927)

The crisis upset every social activity and expression. Direct reference to the object of (artistic or scientific) interest, got into deep crisis, with the very substance of human existence, in the tormented and desperate conception of human existence in Expressionism, as well as in the developments of philosophical thought. The rationality of human behaviour was subverted by Freud's unconscious. All that reflected in the artistic expressions, in the deconstruction of the object; Schönberg upset the rules and concepts of classical harmony.

It is difficult not to see a parallel with the turn in Physics. Paul Forman (1971) has developed an extremely detailed and careful reconstruction of the cultural factors that underlined the choices at the basis of the formulation of Quantum mechanics, making reference also to the changes in the organization and promotion of science in Germany. We think however that a complete analysis should be rooted in a more general consideration of the economic and social contradictions in the Twenties, mainly in Germany (early outlines for such an analysis were laid by Baracca, Livi and Ruffo 1979-80, Donini 1982; very interesting also Rohrlich 1992, von Meyenn 1992, 1994). The tormented history of the Weimar Republic was the melting pot of innovative and contrasting experiences, the clash of rational (Rathenau, the Bauhaus, only as examples) and irrational approaches, which finally resulted in the 1929 crash.

As a matter of fact, *the new generation of physicists who formulated Quantum Mechanics rejected any reference to the description in space and time of the state of the system and its evolution, conceiving the description of atomic systems as a mere* correlation *among their possible states, which could not be further specified*. This point of view turned out to be incompatible with the previous view we have discussed: behind the incompatibility of the physical structures and consequences, the respective authors held since the beginning opposite concepts about what reality is, and what science describes. Heisenberg wrote to Pauli: "The more I weigh up the physical part of Schrödinger's theory, the more horrible it appears to me"; while to

---

[10] About Planck's social positions, one should recall that at the outset of the war he endorsed the 1914 *Manifesto of the Ninety-Three German Intellectuals to the Civilized World* (http://wwi.lib.byu.edu/index.php/Manifesto_of_the_Ninety-Three_German_Intellectuals), although together with other scientists, such as Felix Klein, Haber, Nernst (the two had directed the creation of the German chemical stockpile), Ostwald, Rõntgen, Wien.

Schrödinger, Heisenberg's theory "Made me depressed" (about Schrödinger see Renn 2013). We deem it remarkable that the material and ideological climate in postwar Britain was different (De Maria and La Teana 1982):

> "Dirac was much less disposed than his German colleagues to abandon a spatial-temporal description of microscopic phenomena, and he struggled to construct the new Quantum Mechanics as a *generalization* of (and not only a *break* with) classical physics, through a systematic utilization of the classical Hamilton formalism."

Many considerations could be added on this *new radical turn in physics in the 1920s*, which brought to a completion the Twentieth century "revolution" in atomic physics, but we will not further lengthen in this draft. We will just add some further suggestions related to the more general connections of this turning point, without any aim of systematic nature.

Generally speaking, the problem of controlling the increasing instability at an economic and social level (monetary chaos, inflation, overproduction crisis, social disorder, etc.), which contrasted with the previous view of an inexhaustible pace of economic growth, posed the need of more flexible and effective control tools. The shift from a rigid organization of labour and production towards an organization based on the statistical correlation of an increasing flux of products and goods (including the labour force), led to the adoption of the "Scientific Organization of Labour", aimed precisely at the coordination, correlation and control of multiple, unpredictable factors (Devinat, 1927). The most exact among natural sciences was the most suitable for subsuming and formalizing such goals, and succeeded in incorporating them into a formal structure.

It is interesting to remark how Quantum Mechanics has constituted the basis for subsequent scientific developments. The aftermaths of the 1929 crash reinforced the necessity for a full flexibility of productive and technological innovation. The birth and proliferation of specialized scientific disciplines in the United States in the 1930s supported a continuous differentiation of productive sectors and output, avoiding the accumulation of overproduction crises. In this new situation also the role of scientists and their conceptions and activities changed (one must recall also the brain drain of German scientists): it could no longer consist in providing general representations of the world, but rather the most flexible frame of reference, such that it did not limit, but rather stimulate free developments and practical solutions. Quantum Mechanics had in some sense anticipated and embodied these needs, and presented itself as the ideal framework for these developments: it provided that context and criterion of control and validation on proliferating scientific developments which had been provided in previous phases respectively by mechanism-reductionism, and by the realistic space-time description. It was a *formal* basis instead of a *substantial* or *realistic* one, but it was just what was needed. We deem it very interesting the methodological analogies that have been remarked (Cini 1985, Montagnini 1999-2000) with Norbert Wiener's information theory, in which information too is considered and formalized independently from its specific content and meaning, and the convergence between Wiener's and von Neumann's views.

**References**

Arias Avila Nelson, and Baracca Angelo, 2006, "¿Quién propuso la 'fórmula de Rayleigh-Jeans'?", *Llull, Revista De La Sociedad Espanola De Historia De Las Ciencias Y De Las Tecnicas*, 29, 5-18 (*casanchi.com/fis/rayleigh_jeans01.htm*).

Badino Massimiliano, 2000, The odd couple: Boltzmann, Planck and the application of statistics to physics (1900-1913), *Ann. Phys.*, 18, No 2-3, 81-101.

Badino Massimiliano, 2014, How theories begin: a historical-epistemological study of Planck's black-body radiation theory, *Studies in the History and Philosophy of Science*, Part A, September 25, 2012. pp. 1-39.

Baracca Angelo, 2005, Annus Mirabilis: the roots of the 20[th]-century revolution in physics and the takeoff of the quantum theory, *Llull*, *Revista De La Sociedad Española De História De Las Ciencias Y De Las Tecnicas*, *28*, 295-292 (*dialnet.unirioja.es/descarga/articulo/2470585.pdf*).

Baracca Angelo, Livi Roberto, and Ruffo Stefano, 1979-80, "Le tappe dello sviluppo della teoria dei quanti nel quadro della seconda rivoluzione industriale e delle contraddizioni del capitalismo del primo dopoguerra", *Testi e Contesti:* Part I, *2*, 7-51 (1979); Part II, *3*, 51-80 (1980).

Baracca Angelo, Ruffo Stefano, and Russo Arturo, 1979, *Scienza e Industria 1848-1915*, Bari, Laterza.

Büttner Jochen, Renn Jürgen, and Schemmel Natthias, 2003, Exploring the limits of classical physics: Planck, Einstein, and the structure of a scientific revolution, *Studies in the History and Philosophy ofd Modern Physics*, *34*, 37-59.

Cini Marcello, 1985, "The Context of Discovery and the Context of Validation: The Proposals of Von Neumann and Wiener in the Development of 20th-Century Physics," *Rivista di Storia della Scienza*, *2*: 99-122.

Darrigol O., 1991, Statistics and combinatories in early quantum theory, II: early symptoma of indistinguishability and holism, *Historical Studies in the Physical and Biological Sciences* (pp. 237-298), Berkeley, University of California Press.

Darrigol O., 1992, *From "c"-numbers to "q"-numbers. The classical analogy in the history of quantum theory*, Berkeley, University of California Press.

De Maria Michelangelo, and La Teana F., 1982, Schrödinger's and Dirac's unorthodixy in quantum mechanics, *Fundamenta Scientiae*, *3*: 2, 129-148.

Devinat P., 1927, *L'Organistion Scientifique du Travail en Europe*, Bureau International du Travail, Geneva.

Donini Elisabetta, 1982, *Il Caso dei Quanti*, Milano, Clup-Clued.

Einstein Albert, 1902, Kinetische Theorie der Wärmegleichgewichtes und des zweiten Hauptsatzes der Thermodynamik, *Annalen der Physik*, *9*, 417-433.

Einstein Albert, 1903, Eine Theorie der Grundlagen der Thermodynamik, *Annalen der Physik*, *11*, 170-187.

Einstein Albert, 1904, Zur allgemeinen molekularen Theorie der Wärme, *Annalen der Physik*, *14*, 354-362.

Einstein Albert, 1905a, Über einen die Erzeugung und Verwandlung des Lichts betreffenden heuristichen Gesichtspunkt, *Annalen der Physik, 4, XVII*, 132-148.

Einstein Albert, 1905b, Über die von der mulekularkinetischen Theorie der Wärme geforderte bewegung von in ruhenden Flüssigkeiten suspendierten Teilchen, *Annalen der Physik, 4, XVII*, 549-560.

Einstein Albert, 1905c, Zür Elektrodynamik bewegter Körper, *Annalen der Physik, 4, XVII*, 891-921.

Einstein Albert, 1905d, Ist die Trägheit eines Körpes von seinem Energieinhalt abbängig?, *Annalen der Physik, 4, XVII*, 639-641.

Forman Paul, 1971, Weimar culture, causality and quantum theory 1918-1927: adaptation of German physicists to a hostile intellectual environment, *Historical Studies in the Physical Sciences, 3*, 1-115.

Forman Paul, Heilbron John L, and Weart Spencer, 1975, Physics *circa* 1900: Personnel, Funding, and Productivity of the Academic Establishment, *Historical Studies in the Physical Sciences*, *5*, 1-185.

Galison P., 1981, Kuhn and the quantum controversy, *The British Journal for the Philosophy of Science, 31(1)*, 71-85.

Gearhart Clayton A., 2002, Planck, the quantum, and the historians, *Physics in Perspective*, 4, 170-215.

Gibbs Willard, 1902, *Elementary Principles in Statistical Mechanics Developed with Special reference to the Rational Foundation of Thermodynamics*, Yale University Press; reprinted New York, Dover, 1960.

Heilbron J. L., *The Dilemmas of an Upright Man, Max Planck as Spokesman for German Science*, University of California Press, 1986.

Janik Allan and Toulmin Stephen, *Wittgenstein's Vienna*, London, Weidenfeld and Nicholson, 1973.

Jost M., 1995, *Das Mürchen vom Enfelbeinernen Turm: Reden und Ansätzen*, Lecture Notes in Physics,

*M34, Monographs, Vol. 34*, Berlin, Springer.

Klein Martin J., 1962, Max Planck and the beginning of the quantum theory, *Archive for the History of Exact Sciences, 1*, 459-479.

Klein Martin J., 1963, Planck, entropy and quanta, 1901-1906, *The Natural Philosopher, 1*, 83-108.

Klein Martin J., 1964, Einstein and the wave-particle duality, *The Natural Philosopher, 3*, 3-49.

Klein Martin J., 1966, Thermodynamics and quanta in Planck´s work, *Physics Today, 19*, 23-32.

Klein Martin J., 1967, Thermodynamics in Einstein´s thought, *Science, 157*, 509-516.

Klein Martin J., 1982, Fluctuations and statistical physics in Einstein´s early work, in Holton and Elkana (Eds), *Albert Einstein, Historical and Cultural Perspectives*, Princeton, Princeton University Press, 39-58.

Klein Martin J., Shimeny A., and Pinch T. J., 1979, Paradigme lost? A review symposium, *ISIS, 70 (253)*, 429-440.

Kragh Helge, 2000, Max Planck: the reluctant revolutionary, *Physics World*, December, 31-35.

Kuhn Thomas, 1970, *The Structure of Scientific Revolutions* (2$^{nd}$ edition), Chicago, University of Chicago Press.

Kuhn Thomas, 1978, *Black-body Theory and the Quantum Discontinuity, 1894-1912*, New York, Oxford University Press.

Kuhn Thomas, 1982, *Was sind wissenschaftliche Revolutionen, 10, Werner-Heisenberg-Vorlesung gehalten in München-Nimpherburg am 24 Febr. 1981*, München-Nimpherburg, Carl-Friedrich-von-Siemens-Stiftung.

Kuhn Thomas, 1984, Revisitng Planck, *HSPS, 14(2)*, 232-252.

Landes D. S., 1965, Technological change and industrial development in Western Europe, 1750-1914, in *Cambridge Economic History of Europe*, Vol. VI, *The Industrial Revolution and After*, Cambridge University Press; edited, and extended, as a book, 1969, *The Unbound Prometeus: Technological Change and Industrial Development in Western Europe from 1750 to the Present*, Cambridge University Press.

Montagnini Leone, 1999-2000, *Bit & Plutonium, inc. Le relazioni tra Norbert Wiener e John von Neumann alle origini della cibernetica*, in «Atti dell'Istituto Veneto di Scienze, Lettere e Arti. Classe di Scienze Fisiche, Matematiche e Naturali», *158*, fascicolo II, pp. 361-390.

Needell A. A., 1980, *Irreversibility and the Failure of Classical Dynamics: Max Planck Work on the Quantum Theory, 1900-1915*, Yale University.

Planck Max, 1900a, Entropie und Temperatur strahlender Wärme, *Annalen der Physik, 4*, 719-737.

Planck Max, 1900b, Über eine Verbesserung der Wien´schen Spectralgleichug, *Verhandlungen der Deutschen Physikalische Gesellschaft, 2 (13)*, 202-204.

Planck Max, 1900c, Zur Theorie des Gesetzes der Energieverteilung im Normalspektrum, *Verhandlungen der Deutschen Physikalische Gesellschaft, 2 (17)*, 237-245.

Planck Max, 1901, «Über die Elementarquanta der Materie und Elektrizität», *Annalen der Physik, 4*, 564-566.

Planck Max, 1920, *Die Entstehung und bisherige Entwicklung der Quantentheorie*, Leipzig: Barth, transl. R. Jones and D. H. Williams, *A Survey of Physics*, London, Methuen, 1924.

Planck Max, 1925a, "Zur Frage der Quantelung einatomiger Gase", *Sitz. d. Preuss. Akad. d. Wiss.*, 49-57.

Planck Max, 1925b, "Über die statistische Entropiedefinition", *Berliner Berichte*, 442-451.

Planck Max, 1931, letter to R.W. Wood, October 7, 1931, cited in M.J. Klein, 1966, 27, and Hermann, 1971, *The Genesis of Quantum Theory (1899-1913)*, MIT Press, Cambridge, Massachusetts, 23-24.

Planck Max, 1949, *Wissenschaftliche Selfbiographie*, Leipzig, barth, transl. Franck Gainor, *Scientific Autobiography and Other Papers*, New York, Philosophical Library, 1949.

Rayleigh, Lord, 1900, "Remarks upon the law of complete radiation", *Phil. Mag., 49*, 539-40.

Renn Jürgen, 1993, Einstein as a disciple of Galileo. A comparative study of conceptual development in physics, *Science in Context, 6*, 311-341.

Renn Jürgen, 2006, *Auf den Schultern von Riesen und Zwergen*, Weinheim, Wiley-VCH.

Renn Jürgen (Ed.), 2007, *The Genesis of General Relativity, 4 volumes*, Dordrecht Springer.

Renn Jürgen, 2013, *Schrödinger and the Genesis of Wave Mechanics*, Max-Planck Institute for the History of Science, Berlin, Preprint 437.

Rohrlich F., 1992, "Las interacciones ciencia-sociedad a la luz de la mecánica cuántica y de su interpretación", in L. Navarro (Ed.), *El Siglo de la Física*, Barcelona, Tusquet.

Sommerfeld Arnold, 1927, «Über komische Strahlung», *Südd. Monatshefte*, *24*, 195-98; repr. in Sommerfeld, *Gesammelte Schriften*, Braunshweig, 1968, *4*, 580-83.

von Meyenn K. (Ed.), 1992, "Las interacciones ciencia-sociedad a la luz de la física atómica y subatómica", in L. Navarro (Ed.), *El Siglo de la Física*, Barcelona, Tusquet.

von Meyenn K. (Ed.), 1994, *Quantenmechanik und Weimarer Republik*, Braunschweig, Vieweg.